\documentclass[amsmath, amssymb, reprint, aps, pra]{revtex4-2}

\usepackage{graphicx} 
\usepackage{hyperref} 
\usepackage{mathptmx} 
\usepackage{xcolor}   
\usepackage{dcolumn}  
\usepackage{ulem}     
\usepackage{siunitx}
\usepackage[version=4]{mhchem}
\usepackage{braket}
\usepackage{amsmath}
\usepackage{booktabs}
\DeclareMathAlphabet{\pazocal}{OMS}{zplm}{m}{n} 

\hypersetup{
    colorlinks=true,
    linkcolor=blue,
    citecolor=blue,
    filecolor=blue,
    urlcolor=blue
}

\graphicspath{{./}{./figs/}}

\newcommand{\pref}[2]{\hyperref[#1]{\ref{#1}(#2)}}
\newcommand{\preff}[2]{\hyperref[#1]{\ref{#1}#2}}
\newcommand{\eqpref}[1]{\hyperref[#1]{(\ref{#1})}}

\newcommand{\squig}{{\raise.17ex\hbox{$\scriptstyle\sim$}}}

\newcommand{\cycling}{$\text{a}\,^7\text{S}_3\rightarrow \text{z}\,^7\text{P}^\text{o}_4$ }

\newcommand{\repumpdupper}{$\text{a}\,^5\text{D}_4\rightarrow \text{z}\,^7\text{P}^\text{o}_4$}
\newcommand{\repumpdlower}{$\text{a}\,^5\text{D}_3\rightarrow \text{z}\,^7\text{P}^\text{o}_4$}

\newcommand{\gndstate}{$\text{a}\,^7\text{S}_3$}
\newcommand{\pstate}{$\text{z}\,^7\text{P}^\text{o}_4$}
\newcommand{\dstateupper}{$\text{a}\,^5\text{D}_4$}
\newcommand{\dstatelower}{$\text{a}\,^5\text{D}_3$}

\begin{abstract}
    We have directly loaded a cryogenic beam of molybdenum atoms into a
    magneto-optical trap. By chirping the detuning of the trapping
    lasers, we were able to enhance the number of atoms loaded into the trap
    by more than a factor of two. 
    We optimize the trapped samples for atom number,  temperature, and 
    lifetime by varying the laser-cooling parameters.
    The laser-cooled
    molybdenum atoms are subsequently transferred to a magnetic trap where we achieve vacuum-limited lifetimes of 100\,ms.
    Comparison of magneto-optical trap and magnetic trap lifetimes allow us to extract partial decay rates of the \pstate{}$\rightarrow$ \dstateupper{} and  \pstate{}$\rightarrow$ \dstatelower{} transitions.  We also provide measurements of the isotope shifts for the \cycling{} transition with up to four times better precision than previously reported. Finally, we discuss the prospect
    of applying the chriped magneto-optical trap to produce laser-cooled samples of
    MgF.
\end{abstract}

\begin{document}

\title{Chirped magneto-optical trap of molybdenum}
\author{Sai Naga Manoj Paladugu}\email[Corresponding author: ]{sainagamanoj.paladugu@nist.gov}
\affiliation{Sensor Science Division, National Institute of Standards and Technology,
Gaithersburg, MD 20899}


\author{Nickolas Pilgram}
\affiliation{Sensor Science Division, National Institute of Standards and Technology,
Gaithersburg, MD 20899}

\author{Stephen Eckel}
\affiliation{Sensor Science Division, National Institute of Standards and Technology,
Gaithersburg, MD 20899}

\author{Eric Norrgard}
\affiliation{Sensor Science Division, National Institute of Standards and Technology,
Gaithersburg, MD 20899}
\maketitle

\section{Introduction}
Over the last forty years, magneto-optical traps (MOTs) of dilute atomic and molecular
vapors have proven to be essential starting points for the production of
degenerate quantum gases \cite{anderson1995, davis1995, demarco1999}, quantum simulation of
condensed-matter systems \cite{greiner2002,bernien2017}, quantum metrology \cite{Ludlow2015, DeMille2024,Schlossberger2024},
searches for physics beyond
the standard model \cite{kennedy2020}, and even trace isotope analysis for dating
geological samples \cite{chen1999,yokochi2019}.
In recent years, there has been growing interest in making cold and trapped samples of
transition metal atoms~\cite{Hemmerling2014, yu2022}, of which molybdenum is one.
These atoms have rich electronic structure that make them promising candidates for
applications including novel frequency metrology~\cite{GLOWACKI2022, eustice2023, Bothwell2024}, tests of the time-invariance of fundamental constants~\cite{Nguyen2004, Dzuba2022}, and searches for physics beyond the Standard Model \cite{Figueroa2022}.

One difficulty in laser cooling and trapping many transition and rare-earth metals
is that the initial atomic beam conventionally comes from an oven operating at temperatures above $\SI{1000}{\kelvin}$. 
Thus, the atoms have an initial velocity of up to 1000~m/s, and outgassing from the oven poses a challenge to maintaining ultra-high vacuum conditions \cite{bradley2000,
Eustice2025}. 
However, ablating a target with a high-energy pulsed laser in a cryogenic buffer gas beam (CBGB) cell allows one to produce bright beams of various atomic and molecular species with forward velocities of the order of $\SI{100}{\meter/\second}$ without undue vacuum load \cite{Maxwell2005,Hutzler2012}. 
While slower than a conventional oven, the forward velocity of the CBGB is typically still larger than the capture velocity of a MOT.
Two strategies have been employed to further slow the beam.
First, a second stage may be added to a CBGB to further reduce the forward beam velocity at the expense of some atom flux \cite{Lu2011}, and have been used to directly load MOTs of several rare-earth elements \cite{Hemmerling2014}.
Second, Zeeman slowers can efficiently slow the beam to the capture velocity of the MOT have been used to produce MOTs of K and Cd atoms~\cite{Lasner2021, Padillacastillo2025}.
For many molecular species where Zeeman slowing is impractical, chirped or white-light slowing is a viable alternative  \cite{Barry2012, Truppe2017, Collopy2018, Vilas2022, Zeng2024}.
We recently proposed a third strategy---a frequency chirped MOT---to directly load molecules from a two-stage CBGB into a MOT to eliminate the need for additional slowing lasers~\cite{Rodriguez2023}. 

Here, we demonstrate the first MOT of Mo. 
In Section \ref{sec:apparatus} we provide an overview of the apparatus used in this work. 
In Section \ref{sec:beam} we characterize the two-stage CBGB velocity and flux.  
Section \ref{sec: chirped MOT} details optimal loading of 
 our MOT directly from the two-stage CBGB.
Using a chirped MOT configuration increases the number of atoms by roughly a factor of 2.5. We consider the performance of the chirped Mo MOT within the framework of simple kinematic arguments.
In Section \ref{sec:trap} we characterize properties of Mo atoms trapped in the MOT as well as in a subsequent magnetic trap, as previously demonstrated by Hancox, {\it et al.}~\cite{hancox2005}, and achieve a vacuum-limited lifetime of $\SI{100}{\milli\second}$.
In Section \ref{sec:spectroscopy}, the colder temperatures in the CBGB and MOT allow us to measure higher-accuracy values for isotope shifts, frequencies, and transition rates involving the \pstate{} state of Mo than previously reported \cite{aufmuth1978, WHALING1984, Whaling1988, NIST_ASD}.
Finally, we find favorable prospects for applying the chirped MOT to MgF \cite{Rodriguez2023} in Section \ref{sec: MgF}.

\begin{figure}[b!]
   \centering
   \includegraphics[width=\columnwidth]{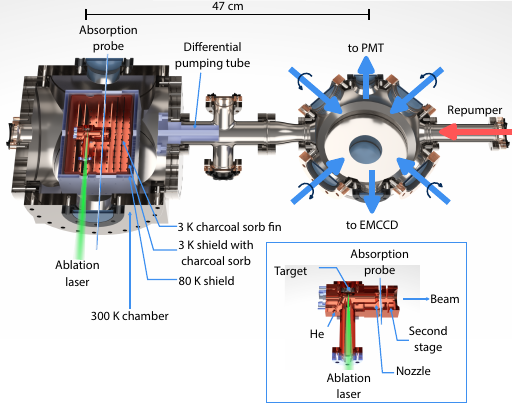}
   \caption{Schematic of the experimental apparatus. The cubic chamber on the left
   houses the CBGB source, and the octagonal chamber on the right houses the MOT.
   Six circular polarized laser beams (two vertical
   beams are not shown) together with two electromagnets in an anti-Helmholtz configuration (not shown) generate the MOT.
   We detect fluorescence from the atomic beam or the MOT via a photomultiplier tube
   (PMT) or an electron multiplying charge-coupled device (EMCCD) camera.
   }
   \label{fig:chamber}
 \end{figure}
 
\section{Apparatus}\label{sec:apparatus}
\begin{figure}[b!]
   \centering
   \includegraphics[width=\columnwidth]{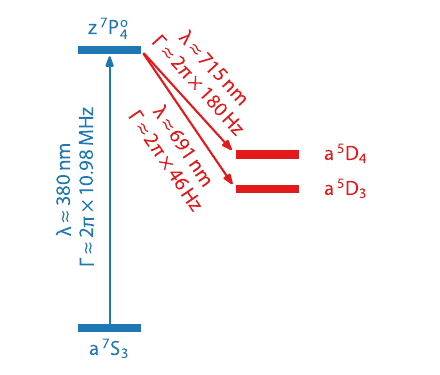}
   \caption{Level diagram with electronic states that are relevant for laser 
   cooling molybdenum. The main cycling transition occurs between the \gndstate{} 
   and the \pstate{} states. From the \pstate{} state, there are leaks to 
   lower-lying \dstateupper{} and \dstatelower{} states.
   We do not show twelve intermediate states in addition to
   $\text{a}\,^5\text{D}_{2}$, $\text{a}\,^5\text{D}_1$, and
   $\text{a}\,^5\text{D}_0$ states; decays to these states
   from $\text{z}\,^7\text{P}_4^\text{o}$ are
   either E1 forbidden, too weak to be observed given our
   statistical precision and
   vacuum-limited lifetime, or forbidden by angular
   momentum and parity selection rules.}
   \label{fig:level_diagram}
 \end{figure}
 
A schematic of the experimental apparatus is shown in Fig.\,\ref{fig:chamber}.
Atoms are ablated from a 99.95\,\% pure Mo disk mounted inside a two-stage CBGB.
Typically, the ablation laser had a pulse energy of $\SI{30}
{\milli\joule}$, a pulse length of $\SI{10}{\nano\second}$, a repetition rate of \SI{1}{\hertz}, and a wavelength of \SI{532}
{\nano\meter}.
The ablation laser was focused to a spot size of approximately $\SI{400}{\micro\meter}$.
Typically, the helium flow rate into the CBGB cell was nominally $q=1.0(1)$ standard cubic centimeters per minute (sccm)~\footnote{Unless stated otherwise, values in parentheses indicate the combined statistical and systematic standard uncertainty}. 
The two-stage CBGB cell was identical 
to the one  previously used to produce MgF in our group \cite{Pilgram2024}, except it
was modified to inject a reactant gas for molecule formation (this feature is not 
utilized in this work).  
The cell was placed inside two layers of thermal radiation shields 
which had approximate temperatures of $\SI{80}{\kelvin}$  and $\SI{3}{\kelvin}$ during 
operation.
Two silicon diodes were used to measure the
cell temperature, which was typically $\SI{3.0(5)}{\kelvin}$ during operation.
The shields were attached to the first and second stages, respectively, of a 
two-stage pulse tube refrigerator rated for $\SI{2.5}{\watt}$ of cooling power at 4\,K.  A helium 
damping pot reduced temperature fluctuations to less than 0.1\,K over the 
pulse tube cycle.

A frequency-doubled Ti:sapphire laser 
system generated the six MOT laser beams driving the \SI{380}{\nano\meter} \cycling{} transition with decay rate $\Gamma = \SI{6.9(5)E7}{\second^{-1}}$ \cite{NIST_ASD} (Fig.\,\ref{fig:level_diagram}). 
Each MOT laser beam had an elliptical profile
with $1/e^2$ major axis $\SI{54\pm 1}{\milli\meter}$ along the molecular beam axis and
minor axis $\SI{10\pm 1}{\milli\meter}$ in the orthogonal direction.  
Each MOT laser beam had a maximum power of 240 mW, corresponding to a maximum saturation parameter $s = 10$ \footnote{We assume a 10\% uncertainty on the saturation parameter.}.  An acousto-optic modulator was used to proportionally vary the power in the laser beams.
A cavity transfer lock stabilized the Ti:sapphire laser to a frequency-stabilized HeNe laser.
During frequency chirps, a combined sample-and-hold and voltage adder circuit was used to sum the latest transfer cavity error signal with the desired frequency ramp.
We verified that laser produces the desired chirp by observing the intended frequency chirp in the beat signal between the MOT laser and a second laser with a fixed frequency.

Two water-cooled anti-Helmholtz electromagnets with an inner diameter of 140\,mm ensure a uniform gradient over the major axis of the laser beams. The magnetic field may be switched 
off in roughly $\SI{100}{\micro\second}$ using an insulated-gate bipolar transistor.
The magnets are able produce axial gradients of up to $B_z^\prime =$$ \SI{3.0}{\milli\tesla/\centi\meter}$, limited by the maximum current rating of the power supply we employ.
Atoms are loaded in the radial direction of the quadrupole where $B^\prime_r =B^\prime_z/2$.

A photomultiplier tube (PMT) and an electron multiplying charge-coupled device (EMCCD) camera were
positioned on opposite sides of the main MOT chamber to collect fluorescence from the atomic
beam and the MOT.
Bandpass interference filters on the PMT and camera restricted the signal to light with wavelengths near 380\,nm. 

\section{Cryogenic Molybdenum Beam} \label{sec:beam}

\begin{figure}[b!]
    \centering
    \includegraphics[width=\columnwidth]{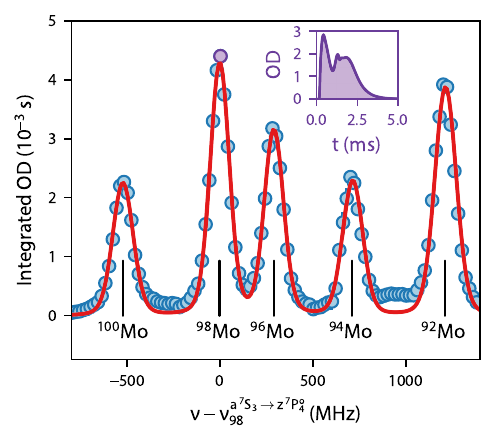}
    \caption{
    Atomic beam absorption as a function of probe frequency $\nu$ referenced to the \ce{^98 Mo} \cycling{} transition frequency $\nu_{98}$.
    The red curve is a fit to the sum of five Voigt profiles.
    The inset depicts a time trace when the probe beam is nearly resonant with the \ce{^98 Mo} \cycling{}
    transition.
    Error bars representing the standard error in the mean of multiple repetitions are typically the size of the points.}
    \label{fig:absorption}
\end{figure}

A roughly $\SI{100}{\micro\watt}$ laser beam is double passed in the 2.45\,mm gap between the first and second stages of the CBGB cell to measure absorption by the atomic beam.
A typical resonant absorption trace as a function of time after the ablation pulse $t$ is shown in the inset to Fig.\,\ref{fig:absorption}.
The majority of the atomic beam exits the first stage of the cell within $\SI{0.5}{\milli\second}$, while a slow tail continues for roughly $\SI{2}{\milli\second}$.
Time-integrated absorption as a function of laser frequency is shown in the main panel of Fig.\,\ref{fig:absorption}.  
Absorption peaks due to the five stable bosonic isotopes of Mo are clearly visible.
A fit to the sum of five Voigt profiles shows deviations from the data between peaks owing to the multitude of hyperfine transitions in the fermionic $^{95}$Mo and $^{97}$Mo isotopes.  
For the most abundant isotope, $^{98}$Mo, the number of atoms produced after the first
stage is roughly $\SI{3E11}{pulse^{-1}}$ and the corresponding beam brightness is roughly 
$\SI{2E12}{sr^{-1}pulse^{-1}}$, comparable to other transition
metal beams produced in CBGB cells \cite{Hutzler2012}.
Additional fermionic structure (not shown) is present out to roughly \SI{3}{\giga\hertz} above the $^{98}$Mo peak. 

\begin{figure}
    \centering
    \includegraphics[width=\columnwidth]{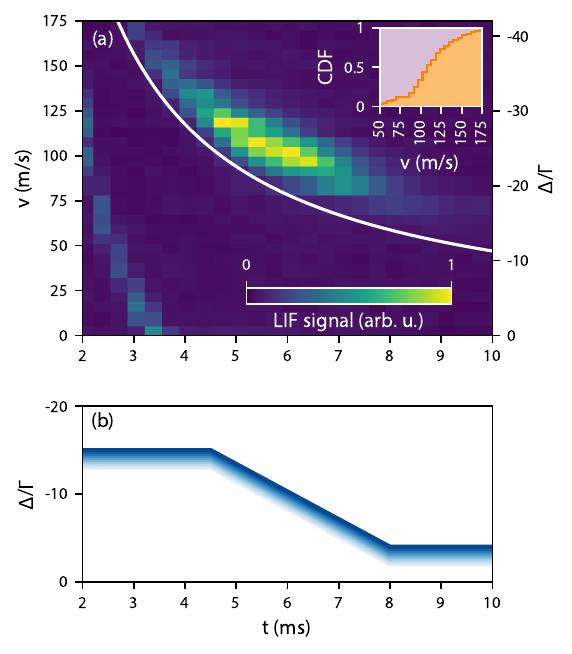}
    \caption{(a) Laser-induced fluorescence (LIF) of the
    \ce{^98 Mo} beam
    in the MOT region as a function of time after the ablation pulse and laser frequency, the latter determining the velocity probed through the Doppler shift.
    The white curve indicates the arrival time of an atom that was created at 
    $t = 0$ with velocity $v$. The inset in the upper right shows the cumulative 
    distribution function (CDF) of velocities in the atomic beam.
    The faint streak on the lower left is likely due to fast atoms of another isotope.
    (b) Exemplary frequency chirp of the MOT beam detuning.}
    \label{fig:velocity_chirp}
\end{figure}

To determine the atomic beam velocity, we alternately apply a single MOT laser beam, either transverse to or oriented at 45 degrees to the molecular beam.   The laser beams are irised to roughly 10\,mm to minimize residual Doppler broadening in this measurement. Using the PMT to detect laser-induced fluorescence (LIF) in these two configurations, we determine the beam velocity from the observed Doppler shift.  The measured atomic beam velocity and time of arrival for
\ce{^98 Mo} are shown in Fig.\,\ref{fig:velocity_chirp}.  The most probable velocity is found to be 105\,m/s, and a full width at half maximum (FWHM) of roughly 40\,m/s.

From the LIF and total detection efficiency of the PMT, the atomic flux in the MOT region is approximately $\SI{1E6}{pulse^{-1}}$.
Accounting for line-of-sight losses to the MOT region, there is an apparent factor of 1000 reduction in flux between the CBGB first stage cell and the PMT region.
The typical extraction efficiency of a second stage of a two-stage CBGB has been measured to be roughly 1\,\% \cite{Lu2011}.
The excess loss by a factor of 10 in our system is currently being investigated.

\section{Chirped Molybdenum MOT} \label{sec: chirped MOT}

\begin{figure*}
    \centering
    \includegraphics[width=\textwidth]{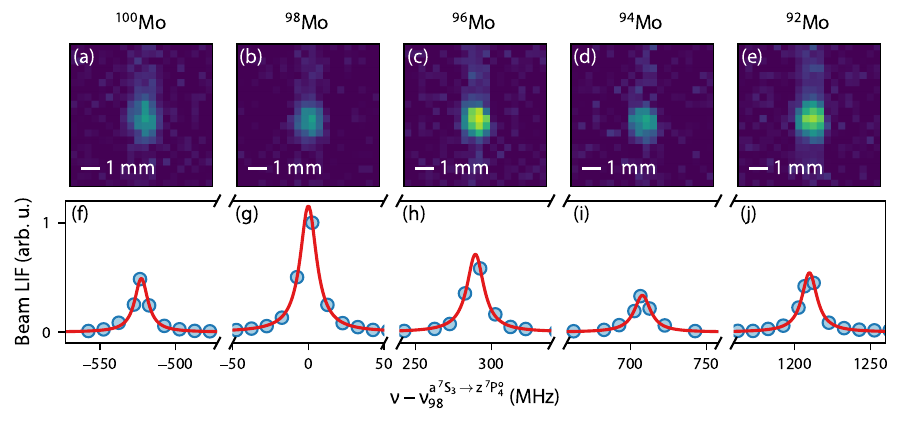}
    \caption{(a--e) MOT images for all stable bosonic isotopes of molybdenum.
    Here, the final MOT detuning was $\Delta_f = -2\Gamma$, the six beam
    saturation parameter was  $s = 10$, and the axial gradient was
    $B'_z=\SI{3}{\milli\tesla/\centi\meter}$, and no repump laser is present. The numbers of atoms in each image
    from left to right are approximately $3\times 10^3$, $4\times 10^3$,
    $5\times 10^3$, $3\times 10^3$, and $5\times 10^3$.
    (f--j) LIF signal from molybdenum beams detected in the MOT region as a
    function of detuning from the \ce{^98 Mo} \cycling{} transition.}
    \label{fig:isotopes}
\end{figure*}

\begin{figure}
    \centering
    \includegraphics[width=\columnwidth]{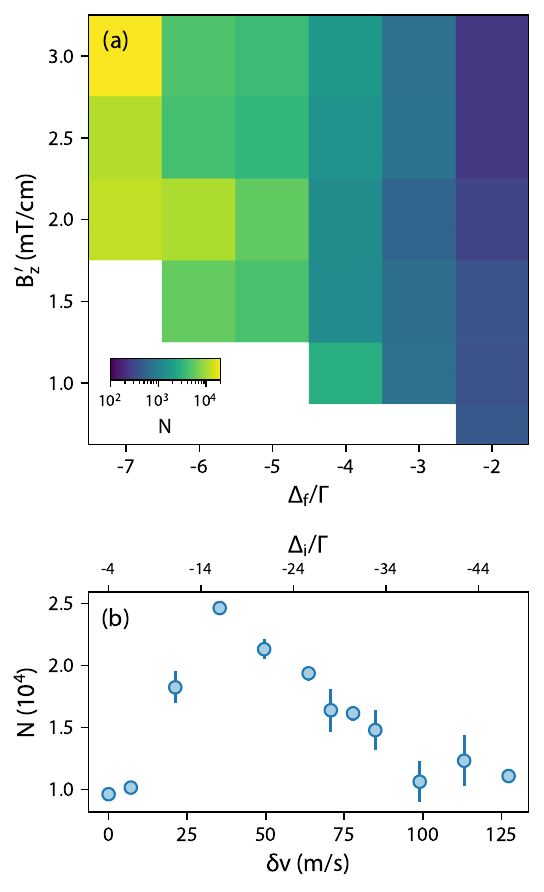}
    \caption{(a) Optimization of the number of trapped atoms $N$ in the MOT as a 
    function of the axial gradient $B^\prime_z$ and
    final detuning of the MOT beams $\Delta_f$. In all cases shown here,  the chirp amplitude was $\delta v = \SI{100}{\meter/\second}$
    and the 
    total six beam saturation parameter was $s=10$.  (b) Optimization of $N$ as a function of $\delta v$, with $B^\prime_z = \SI{3}{\milli\tesla/\centi\meter}$, 
     $\Delta_f = -4\Gamma$, and $s=10$. The initial detuning $\Delta_i = \Delta_f
     - k\delta v/\cos\theta$ is shown on the top axis.}
    \label{fig:optimization}
\end{figure}

We observed a MOT of several thousand $^{98}$Mo atoms directly loaded from the CBGB with a static detuning of roughly $\Delta = -4\Gamma$, saturation parameter $s=10$,
and axial gradient $B_z'=\SI{3.0}{\milli\tesla/\centi\meter}$
\footnote{The atom number is calculated from fluorescence measurements using 
$N_\text{atom}=N_\text{photon}/CR_\text{sc}t_\text{exposure}$, where $C$ is the geometric
collection efficiency, $R_\text{sc}$ is the scattering rate, and $t_\text{exposure}$ is
the exposure time. We estimate $C\approx \SI{220E-6}{}$. During imaging, we typically
set $s=10$ and use the final detuning of the MOT lasers to determine $R_\text{sc}$ via
Eq.\,\ref{eq: scattering rate}. Typically, $t_\text{exposure}=\SI{1}{\milli\second}$.}.
Similar to Cr \cite{bradley2000}, we were able to observe short-lived 
($\Gamma^{\rm{MOT}} > \SI{100}{\second^{-1}}$) MOTs without any repumping lasers.  
The excited state of the laser cooling transition \pstate{} may weakly decay to the 
metastable a$^5$D manifold as depicted in Fig.\,\ref{fig:level_diagram}.  To repump the 
\dstateupper{} level, a second Ti:sapphire laser beam intersected 
the MOT counter-propagating to the atomic beam.
This repump laser beam had $\SI{100} {\milli\watt}$ of power in a $\SI{25}{\milli\meter}$ $1/e^2$ diameter.
With the addition of this laser, the MOT decay rate ($\Gamma^{\rm{MOT}}<\SI{25}{\second}^{-1}$) and trapped atom number ($N \leq 1.0\times 10^4$) were substantially improved. 
The $\SI{691}{\nano\meter}$ \repumpdlower{} transition was outside the tuning range of our Ti:sapphire lasers, so we were unable to repump out of this level.

We employed a chirped MOT to enhance loading \cite{Rodriguez2023}.  During the MOT 
loading phase, the laser detuning is ramped linearly from initial detuning $\Delta_i$ to 
final detuning $\Delta_f$ at a chirp rate $\chi = (k/\cos\theta)\delta v/\delta t$, 
where $\delta v = \cos\theta(\Delta_f-\Delta_i)/k$ is the
chirp amplitude in units of 
velocity, $\delta t$ is the chirp duration, $k=|\mathbf{k}|$ is the magnitude 
of the wavevector of the MOT lasers, and $\theta$ is the relative angle between 
$\mathbf{k}$ and the atomic velocity $\mathbf{v}$.
By varying the timing of the chirp, we found the number of captured atoms $N$  was maximized when the chirp began $\SI{4.5}{\milli\second}$ 
after the ablation pulse and lasted for $ \delta t = \SI{3.5}{\milli\second}$.  This empirically determined chirp profile, shown in Fig.\,\ref{fig:velocity_chirp}(b), coincides with the interval when the majority of the atomic beam arrives in the MOT region, shown in Fig.\,\ref{fig:velocity_chirp}(a).  Suboptimal chirp delay or duration (by several milliseconds) was found to trap fewer atoms than the static MOT.

Without the repump laser present, we observed chirped MOTs of
all five stable bosonic Mo isotopes [Fig.\,\ref{fig:isotopes}(a-e)] 
with $s = 10$, $B'_z = \SI{3.0}{\milli\tesla/\centi\meter}$,
$\delta v = 100\,\si{\meter/\second}$, and $\Delta_f = -2\Gamma$ from the relevant field-free resonance observed in the atomic beam LIF [Fig.\,\ref{fig:isotopes}(f-j)].
Several thousand atoms of each isotope were trapped in this manner, crudely in 
proportion to their natural isotopic abundance. 
The remainder of this work focuses on trapping the most abundant isotope, $^{98}$Mo.

We optimized the captured atom number $N$ as a function of $\Delta_f$ and
the axial gradient $B_z^\prime$ shown in Fig.\,\ref{fig:optimization}(a). Here, we 
set $s = 10$ and $\delta v = \SI{100}{\meter/\second}$.
The most atoms are captured when using the largest gradient that we can apply, $B_z^\prime =$$ \SI{3.0}{\milli\tesla/\centi\meter}$.
The trapped atom number also increases with more negative $\Delta_f$.
However, for $\Delta_f < -7\Gamma$, the MOT cloud has such a large spatial extent that an accurate atom number is difficult to extract from a Gaussian fit to the camera image. Thus, we typically performed
chirped loading into a MOT that has $\Delta_f = -7\Gamma$, $B_z^\prime =\SI{3.0}{\milli\tesla/\centi\meter}$, and a total six-beam saturation
parameter of roughly $s=10$. 
When studying the MOT in another trap condition
to characterize the temperature or lifetime, we subsequently
ramp the laser detuning and/or power over a $\SI{10}{\milli\second}$ interval after the initial chirped load.

Finally, we optimized the number of atoms $N$ loaded in the chirped MOT  as a function of chirp amplitude $\delta v$.  Fig.\,\ref{fig:optimization}(b) shows the number of atoms loaded into a MOT with $s=10$,  $B_z^\prime = \SI{3.0}{\milli\tesla/\centi\meter}$, and $\Delta_f = -4\Gamma$. We found that the optimum chirp amplitude of $\delta v=\SI{35}{\meter/\second}$ increases the atom number to roughly $N=2.4\times 10^4$, which was the largest number we were able to trap
overall.

Our observed increase in $N$ for the chirped MOT over the static MOT can be explained by the chirp compensating for our technically limited magnetic field gradient. 
Consider loading a MOT with laser beam diameter $d$ where the atom with mass $M$ is traveling with an initial velocity $v$ along the radial direction $\mathbf{e}_r$.
The limit to the acceleration assuming constant scattering rate $R_{\rm{sc}}$ is given by $
    a^{\rm{sc}}_{\rm{max}} = R_{\rm{sc}} \hbar k \cos\theta/M$,
where $\theta$ is the relative angle between $\textbf{v}$ and $\textbf{k}$
and $\hbar$ is the reduced Planck constant.
From simple kinematic arguments \cite{HAUBRICH1993, Tiecke20009, Lunden2020}, it may be shown that scattering-rate-limited capture velocity is
\begin{equation}\label{eq:capture velocity scatter}
v_{\rm{c}}^{\rm{sc}} = \sqrt{2dR_{\rm{sc}} \hbar k /M },
\end{equation}
while equating the Doppler shift to the sum of the Zeeman shift and laser detuning 
$\Delta$ at the edge of the laser beam, $d/2 \cos\theta$, determines the the gradient 
limit to the capture velocity
\begin{equation} \label{eq: capture velocity gradient}
   v_{\rm{c}}^{\rm{grad}}  =  \frac{g^*\mu_B }{2\hbar k \cos^2 \theta} B^\prime_r d - \frac{\Delta}{k\cos\theta}.
\end{equation}
For the present work, $\theta = \pi/4$, $d \approx \SI{54}{\milli\meter}$ is the major-axis of the MOT beams, the technically limited gradient was $B^\prime_r = \SI{1.5}{\milli\tesla/\cm}$, the laser detuning was $\Delta=-4\Gamma$, and $g^* = 1.992$ is the largest differential $g$-factor of Zeeman sublevels in the cycling transition.
We estimated that $R_{\rm{sc}} \approx 0.2 \Gamma$ during MOT loading by performing a three-dimensional rate equation model using {\tt pylcp}~\cite{Eckel2022}.
Thus, we estimate that under static conditions, our capture velocity $v_{\rm c}$ is gradient limited with $v_{\rm{c}} =  v_{\rm{c}}^{\rm{grad}} \approx \SI{110}{\meter/\second}$.
We observed an optimal number of atoms when chirping with $\delta v =\SI{35}{\meter/\second}$, such that $v_{\rm{c}} =  v_{\rm{c}}^{\rm{grad}} + \delta v \approx\SI{145}{\meter/\second}$.
The optimal chirp capture velocity is crudely in line with our estimated scattering rate-limited capture velocity $v_{\rm{c}}^{\rm{sc}}  \approx\SI{125}{\meter/\second}$.  
From the cumulative distribution function (CDF) of the observed velocity profile 
shown in the upper right inset of Fig.\,\ref{fig:velocity_chirp}(a), we estimate a  1.8 
times increase in $N$ when  $v_{\rm{c}}$ is increased from $v_{\rm{c}}^{\rm{grad}}$ to 
$v_{\rm{c}}^{\rm{sc}}$, roughly in line with our observed 2.4 times increase when chirping.

\section{Molybdenum Trap Properties} \label{sec:trap}

\begin{figure}
    \centering
    \includegraphics[width=\columnwidth]{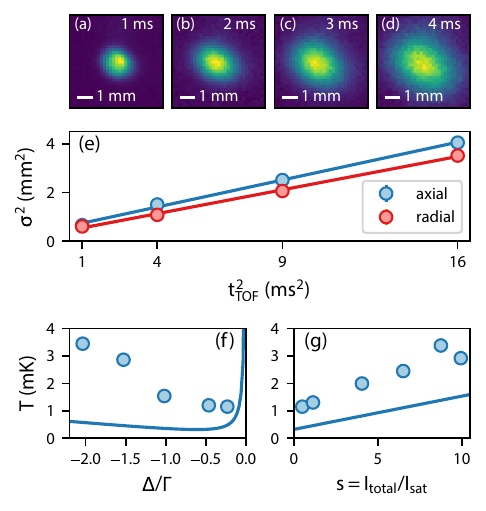}
    \caption{Temperature measurements (a-d) Time of flight images
    of  \ce{^98 Mo} atoms released from a MOT with a final detuning of $-\Gamma/2$ and a total
    six beam saturation parameter of $s = 0.45$. (e) $1/e$ radii $\sigma$, extracted from the images from (a-d), as a function of time of flight $t_{\rm{TOF}}$. The lines are linear fits to the data. (f)
    Temperature $T$ as a function of MOT detuning $\Delta$ with six beam saturation parameter $s=0.45$. (g)
    $T$ as a function of $s$ with
    $\Delta = -\Gamma/2$. The solid blue curves in (f,g) are the predictions
    from the one-dimensional Doppler cooling model Eq.\,\ref{eq:1D Doppler}.}
    \label{fig:tof}
\end{figure}

The temperature of atoms in the MOT was determined by the ballistic expansion method \cite{Lett1988}.
We synchronously switched off the magnetic field and trapping light for a variable time 
of flight, $t_{\rm{TOF}}$. The tapping light was then turned back on, and the 
the EMCCD camera imaged the ballistically expanded atomic cloud for $\SI{1}{\milli\second}$ 
[Fig.\,\ref{fig:tof}(a-d)].
Fitting the axial and radial $1/e$ radii $\sigma_{z}$ and $\sigma_r$ as a function of $t_{\rm{TOF}}$ yields the temperatures $T_z$ and $T_r$ in the respective dimensions [Fig.\,\ref{fig:tof}(e)].  
In Fig.\,\ref{fig:tof}(f), the circles show the geometric mean temperature $T = T_r^{2/3}T_z^{1/3}$ as a function of MOT laser detuning $\Delta$ with $s=0.45$. A heuristic one-dimensional Doppler cooling model \cite{Lett1989} predicts the temperature to be
\begin{equation}\label{eq:1D Doppler}
    T(s,\Delta) = - \frac{\hbar \Gamma^2}{8k_{\rm{B}} \Delta} \left(1+s+4\frac{\Delta^2}{\Gamma^2}\right).
\end{equation}
Here, $\hbar$ is the reduced Planck constant and $k_{\rm{B}}$ is the Boltzmann constant.
According to Eq.\,\ref{eq:1D Doppler}, $T$ decreases monotonically as $s\rightarrow 0$, and the minimum temperature $T_{\rm{D}} = \hbar\Gamma/2k_{\rm{B}}\approx \SI{260}{\micro\kelvin}$ is achieved when $s\ll 1$ and $\Delta = -\Gamma/2$.  
The one-dimensional Doppler cooling model of Eq.\,\ref{eq:1D Doppler} is shown in Fig.\,\ref{fig:tof}(f,g) as a solid line. 
The observed temperature decreases with more positive detuning and lower intensity as expected, but is found to be several times higher than the heuristic model predicts.  
The lowest temperature observed is $T = \SI{1.1}{\milli\kelvin}$, or approximately $4T_{\rm{D}}$.
  
We note that the level structure of Mo is similar to that of Cr and should be amenable to  sub-Doppler cooling methods such as gray molasses \cite{Gabardos2019} and blue-detuned magneto-optical trapping \cite{Jarvis2018}.  
While we made no effort at sub-Doppler cooling in this work, it should be possible in the future to cool Mo using one or more of the a$^7$S$_3 \rightarrow $z$^7$P$^\circ_{2,3,4}$ transitions to a few times the recoil temperature, $T_{\rm{R}} = (\hbar k)^2/2Mk_{\rm{B}}\approx\SI{1.4}{\micro\kelvin}$, where $M$ is the atomic mass.

\begin{figure*}
    \centering
    \includegraphics[width=\textwidth]{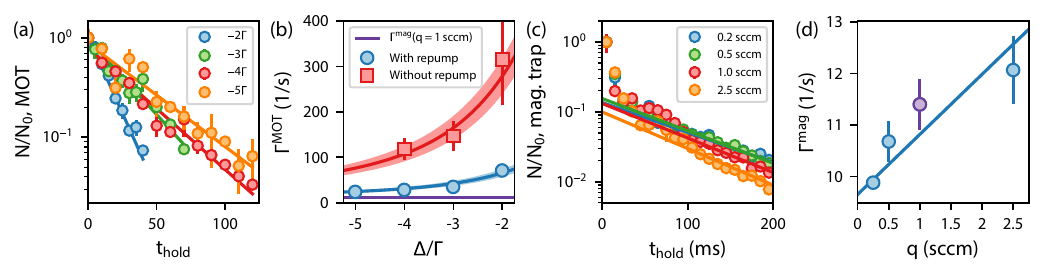}
    \caption{(a) Number of trapped atoms $N$ in the MOT vs. time $t_{\rm hold}$ for various detunings.
    Here, $t_{\rm hold}=0$ corresponds to $\SI{30}{\milli\second}$ after the ablation pulse, and we normalize all data to the number $N_0$ at $t_{\rm hold}=0$.
    Lines show exponential fits to extract the decay rate $\Gamma^{\rm MOT}$.
    (b) $\Gamma^{\rm MOT}$ as a function of laser detuning $\Delta$.
    Curves and their associated uncertainty bands are fits to Eq.~\ref{eq:mot loss model}.
    The purple, horizontal line indicates the magnetic trap lifetime $\Gamma^{\rm mag}(q=1.0\mbox{ sccm})$, as extracted from (c).
    (c) $N$ vs. $t_{\rm hold}$ in a magnetic trap for various helium
    flow rates, $q$. Lines show exponential fits to extract the decay rate $\Gamma^{\rm mag}$.
    (d) $\Gamma^{\rm mag}$ vs. $q$.
    The line is a fit to the data.
    }
    \label{fig:lifetime}
\end{figure*}

In Fig.\,\ref{fig:lifetime}, we show the decay rate of the trapped atoms under various conditions. Fig.\,\ref{fig:lifetime}(a) shows $N$ as a function of time $t_{\rm hold} = t - \SI{30}{\milli\second}$ for various detunings $\Delta$, normalized to the atom number $N_0$ observed at $t_{\rm hold}=0$.
The data in Fig.\,\ref{fig:lifetime}(a) were taken under optimal loading conditions: $B_z^\prime =$$ \SI{3.0}{\milli\tesla/\centi\meter}$, $s=10$, and the 715\,nm repump laser present.
The number of atoms in the MOT was found to be well described by a single exponential decay rate $\Gamma^{\rm{MOT}}$.  In Fig.\,\ref{fig:lifetime}(b) we plot the best fit $\Gamma^{\rm{MOT}}$ as a function of $\Delta$ with and without the 715\,nm repump laser present.

By turning off the MOT laser beams after loading, the highly magnetic Mo atoms are readily trapped in the MOT quadrupole field. 
In  Fig.\,\ref{fig:lifetime}(c) we show the number of atoms $N$ remaining in the magnetic trap as a function of time, normalized to the atom number $N_0$ observed $\SI{30}{\milli\second}$ after the ablation pulse.  
No effort was made to optically pump the atoms in the MOT into a low-field seeking state before magnetically trapping.
Thus we observe a fast decay for the first few milliseconds due to loss of atoms in untrapped Zeeman sub-levels. 
After about $\SI{20}{\milli\second}$, approximately 20\,\% of the atoms remain, which are subsequently lost exponentially at a rate $\Gamma^{\rm{mag}}$. 
We found that $\Gamma^{\rm{mag}}(q)$ was weakly modified (about 20\,\%) by varying the helium flow rate $q$ into the CBGB cell by a factor of 10, shown in Fig.~\ref{fig:lifetime}(d).
Extrapolating to a helium flow rate of $q=0$ and assuming the loss rate in the conservative magnetic trap is due entirely to collisions with background gas molecules, we find a vacuum limited loss rate of $\Gamma^{\rm{mag}}(q=0) = \SI{9.7(2)}{\second^{-1}}$.
This loss rate is typical for experiments where the vacuum chamber is not baked to remove excess water, as was the case here.
For large detunings, we observe $\Gamma^{\rm MOT} \rightarrow \Gamma^{\rm{mag}}(q=1\mbox{ sccm})=\SI{11.4(5)}{\per\second}$, as expected and shown in Fig.~\ref{fig:lifetime}(b),
whereas for small detunings the lifetime is limited by leakage
into $\text{a}\,^5\text{D}_{4}$ and $\text{a}\,^5\text{D}_3$ (without 715 nm repump) or
just $\text{a}\,^5\text{D}_{3}$ (with 715 nm repump).

\section{Molybdenum Spectroscopy} \label{sec:spectroscopy}

Using the observed decay rates in the MOT and magnetic trap, along with the previously determined radiative decay rate of the \pstate{} state $\Gamma = \SI{6.9(5)E7}{\second^{-1}}$ \cite{WHALING1984,NIST_ASD}, we were able to determine the partial decay rates $\Gamma_{ij}$ from the excited \pstate{} state to the \dstateupper{} and \dstatelower{} states.  
Our results for the partial decay rates and transition frequencies for $^{98}$Mo are shown in Table \ref{tab:decay rate comparison}.
To obtain the partial decay rates, we assume that the photon scattering rate is
generically
\begin{equation}\label{eq: scattering rate}
    R_{\rm{sc}} = \frac{n_e}{n_g + n_e}\Gamma\frac{s_\text{eff}}{1+s_\text{eff}+(2\Delta/\Gamma)^2},
\end{equation}
where $\Gamma = \Gamma_{^7\rm{P}_4,^7\rm{S}_3}+\Gamma_{^7\rm{P}_4,^5\rm{D}_3} +\Gamma_{^7\rm{P}_4,^5\rm{D}_4}$ and $s_\text{eff} = (n_g + n_e)/2n_g^2 \times s$
is the effective multi-level saturation parameter \cite{tarbutt2013, norrgard2016}.
From our three-dimensional rate equation model of the MOT using
{\tt pylcp} \cite{Eckel2022}, the trapped atoms primarily occupy the stretched angular momentum states $\ket{J=3,m_J=3}$ and $\ket{J'=4,m_J'=4}$.
Thus, we set $n_g = n_e = 1$. The MOT loss rate is 
\begin{equation}\label{eq:mot loss model}
  \Gamma^{\rm{MOT}} = \Gamma^{\rm{mag}}+  R_{\rm{sc}} \frac{\Gamma_{\rm{dark}}}{\Gamma}  
\end{equation}
where $\Gamma_{\rm{dark}} = \Gamma_{^7\rm{P}_4,^5\rm{D}_3} +\Gamma_{^7\rm{P}_4,^5\rm{D}_4}$ without the repump laser present, or $\Gamma_{\rm{dark}} = \Gamma_{^7\rm{P}_4,^5\rm{D}_3}$ when the repump laser is present.
Our decay rate model does not capture potential decays
from \dstateupper{} and \dstatelower{} to \gndstate{} and lower $\text{a}\,^5\text{D}_J$ levels, as these
transitions are not E1 allowed and are
suspected to have rates many orders of magnitude
less than the MOT decay rate \footnote{For isoelectronic \ce{Cr}, the radiative lifetimes of the \dstateupper{} and
\dstatelower{} states are calculated in Ref. \cite{GLOWACKI2022} to be on the order of $\SI{1000}{\second}$.}.

Figure\,\ref{fig:lifetime}(b) shows fits  of $\Gamma^{\rm{MOT}}$ to Eq.\,\ref{eq:mot loss model} for the case with and without the repump laser present. In both cases, we fix $s = 10$,
$n_g = n_e = 1$, $\Gamma = 2\pi\times\SI{10.98}{\mega\hertz}$, and $\Gamma^\text{mag} =
\SI{11.4}{\per\second}$.
With the repump laser present,  the fit yields $\Gamma_{\rm{dark}} = \Gamma_{^7\rm{P}_4,^5\rm{D}_3} = \SI{290(30)}{\second}^{-1}$.
Without the repump laser present, $\Gamma_{\rm{dark}} =  \SI{1440(200)}{\second^{-1}}$, and we find that  $\Gamma_{^7\rm{P}_4,^5\rm{D}_4} =\SI{1150(200)}{\second}^{-1}$. The errors account for 10\% uncertainty in $s$, a 7\%
uncertainty in the total decay rate $\Gamma$, and the statistical
uncertainty from the fit. The possibility of additional, two-body loss mechanisms is ruled out by the good statistical agreement of the decay rate measurement with a single exponential model in all cases. While we cannot
rule out decays to $\text{b}\,^5\text{D}_{J}$ and $\text{c}\,^5\text{D}_{J}$ levels, 
simple scaling arguments show that the decay rates to these states are smaller than our statistical uncertainties \cite{NIST_ASD}. Thus, we do not report them here.
\begin{table}[t]
\begin{tabular}{l S[table-number-alignment = right] S[table-number-alignment = right] cc} \toprule
Parameter
& \text{\cycling{}}
& \text{\repumpdupper{}}
& \text{\repumpdlower{}} 
& Reference\\ \midrule
$\nu_{ij}$ (GHz) & 789066.6 & 418934.0 & 433557.3\textsuperscript{*} & \cite{NIST_ASD}   \\   
                 & 789065.93(55) & 418933.67(28) & -- & this work   \\
$\Gamma_{ij}$ (\SI{}{\second}$^{-1}$)  & \multicolumn{1}{c}{ 6.9(5) $\times 10^7$ }&\multicolumn{1}{c}{ 6(2)$\times 10^5$ } & --  & \cite{NIST_ASD}\\
                        &    \multicolumn{1}{c}{ 6.9(2) $\times 10^7$ } &        \multicolumn{1}{c}{ -- }          &    \multicolumn{1}{c}{ -- }      &\cite{WHALING1984}\\\
                        &  \multicolumn{1}{c}{ -- }  & \multicolumn{1}{c}{ 4 $\times 10^4$ }     & \multicolumn{1}{c}{ -- }  & \cite{Palmeri1998}\\

                         & \multicolumn{1}{c}{ -- } & \multicolumn{1}{c}{1.15(20)$\times 10^3$ }& \multicolumn{1}{c}{2.9(3)$\times 10^2$}& this work\\
\bottomrule
\end{tabular} \caption{Comparison of our measured transition frequencies and decay rates to values from previous works. Values in parentheses are the combined $1\sigma$ statistical and systematic uncertainty.  Values mark with an asterisk are inferred but not explicitly provided in Ref.\cite{NIST_ASD}.  Values for this work are for $^{98}$Mo.}
\label{tab:decay rate comparison}
\end{table}

To the best of our knowledge, this work is the first reported value for the weak transition rate $\Gamma_{^7\rm{P}_4,^5\rm{D}_3}$, and we have substantially reduced the experimental uncertainty in the value of $\Gamma_{^7\rm{P}_4,^5\rm{D}_4}$.
We note that there has been disagreement in the literature regarding the value of $\Gamma_{^7\rm{P}_4,^5\rm{D}_4}$.
Whaling and Brault measured the transition rates and branching fractions for 2835 transitions of Mo using both a hollow cathode source and an inductively coupled argon plasma source \cite{Whaling1988}. 
In Appendix A of Ref.\,\cite{Whaling1988}, they noted that the z$^7$P$_{2,3,4}^\circ$ branching fractions to the ground state are all approximately 98\,\%, but strong self-absorption of these lines in the hollow cathode source artificially decreased their observed branching ratios of these transitions to the the various z$^7$P$^\circ \rightarrow $ a$^5$D transitions. Surprisingly, the reported transition rate in Ref.\,\cite{Whaling1988} is given only a 3\,\% uncertainty  despite noting this systematic effect.
Palmeri and Wyart \cite{Palmeri1998} calculated the transition rate using pseudo-relativistic Hartree-Fock calculations and fit the eigenvalues to the observed energies in Ref.\,\cite{Whaling1988}.  While Palmeri and Wyart's value is substantially smaller than the value reported in Ref.\,\cite{Whaling1988}, it still exceeds our value by a factor of 40.

Figure \ref{fig:isotopes}(f-j) shows the LIF of the atomic beam as a function of laser 
frequency near  the \cycling{} transition for each bosonic 
isotope of Mo.  For this measurement, a pair of laser beams irised to a
diameter of $\SI{1.0(1)}{\centi\meter}$ was directed transverse to the atomic beam.
We extracted the isotope shifts $\nu_{A-A^\prime} =\nu_{A} - \nu_{A^\prime}$, where 
$A$ and $A^\prime$ are the atomic mass numbers, and compare to previously measured 
values in Table \ref{tab:isotope_shifts}. The spectra are well characterized by a 
Lorentzian line shape, with all fitted widths found to be within 20\,\% of 
the \pstate{} radiative decay rate $\Gamma/2\pi = \SI{11.0\pm0.8}{\mega\hertz}$. 
The fit uncertainty for each line center was less than $\SI{200}
{\kilo\hertz}$.  The uncertainty in the isotope shifts are primarily due in 
roughly equal parts to nonlinearity and lock point uncertainty in the transfer 
cavity lock used to stabilize the Ti:sapphire laser \cite{Norrgard2022}. Two of our four isotope shift values disagree by more than $2\sigma$ with those reported in Ref.\,\cite{aufmuth1978}. The uncertainties reported in Ref.\,\cite{aufmuth1978} represent only the standard deviation in their data and a discussion of potential systematic errors is not presented. It may be interesting to
perform a more complete King plot analysis, extract the nuclear charge radii,
and compare with those determined in Ref.\,\cite{CHARLWOOD200923} in future work.

The \cycling{} transition isotope shifts in Table \ref{tab:isotope_shifts} and transition frequency $\nu_{ij}$ for $^{98}$Mo in Table \ref{tab:decay rate comparison} may be used to determine the transition frequencies for the other bosonic isotopes.  For the \dstateupper$\rightarrow$\pstate{} transition, our measured transition frequency of $^{98}$Mo may be combined with the isotopes shifts measured by Olsson \textit{et al.} \cite{Olsson1984}.

\begin{table}[]
    \centering
    \begin{tabular}{l D{.}{.}{1} D{.}{.}{-1}} \toprule
        Isotope Shift (MHz) & \multicolumn{1}{c}{Ref.\,\cite{aufmuth1978}} & \multicolumn{1}{r}{This work} \\ \midrule
         $\nu_{92-96}$  & 899(3)  &  920.4(1.9) \\
        
         $\nu_{94-96}$ & 417(5) &  418.4(1.6) \\
        
         $\nu_{98-96}$  & -303(8) & -289.4(1.6) \\
        
         $\nu_{100-96}$ & -803(1) & -812.0(1.9) \\ \bottomrule
    \end{tabular}
    \caption{Isotope shifts  $\nu_{A-A^\prime} = \nu_{A} - \nu_{A^\prime}$ for the Mo 
    \cycling{} transition for all stable bosonic isotopes. All shifts are relative
    to the line center for $\ce{^96Mo}$.  Values in parentheses for our
    work are the combined $1\sigma$ statistical and systematic uncertainties.  Values in parentheses for Ref.\,\cite{aufmuth1978} are the $1\sigma$ statistical uncertainties only.}
    \label{tab:isotope_shifts}
\end{table}

\section{Prospects for Magnesium Monofluoride Laser Cooling} \label{sec: MgF}

Excepting the recent demonstration of laser cooling of AlF~\cite{padillacastillo2025magnetoopticaltrappingaluminummonofluoride}, all magneto-optically trapped molecules have optically cycled using a $^2\Pi_{1/2}$ excited state.  The $g$ factor of  $^2\Pi_{1/2}$ states is small ($g \lesssim 0.1$), such that an applied magnetic field gradient provides little spatial variation to the velocity-dependent force.  Essentially, this restricts the capture velocity of such systems to one given by Eq.\,\ref{eq: capture velocity gradient}.  Recently, we measured vibrational branching fractions \cite{Norrgard2023} and demonstrated optical cycling in MgF using the $A^2\Pi_{1/2}$ excited state \cite{Pilgram2024}, and we proposed a chirped-MOT for laser cooling and trapping of MgF \cite{Rodriguez2023}.

Our results and analysis of the Mo chirped MOT suggests that a chirped MOT of MgF should perform favorably.   Relative to Mo, MgF has a favorable mass $m_{\rm{Mo}}/m_{\rm{MgF}}  = 2.28$,
radiative decay rate $\Gamma_{\rm{MgF}}/\Gamma_{\rm{Mo}} = 1.94$, and wavevector $k_{\rm{MgF}}/k_{\rm{Mo}} = 1.05$.
Thus, the maximum radiative acceleration which can be applied to MgF is roughly 4.6 times larger than Mo.
Through rate equation simulations, we find that the maximum slowing force for our Mo MOT is roughly $0.2\times \hbar k_{\rm{Mo}}\Gamma_{\rm{Mo}}$.  
For MgF, this is reduced to $0.05\times \hbar k_{\rm{MgF}}\Gamma_{\rm{MgF}}$ owing to the type-II level structure used for optical cycling \cite{Rodriguez2023}.  Thus, the net effect of all optical cycling transition parameters predicts an approximately equal radiative acceleration for MgF and Mo.   
The maximum chirp rate, determined by the maximum radiative acceleration, is therefore roughly equivalent  for MgF and Mo.
The chirp rate required to address a given $ \delta v$ is only about 5\% higher MgF due to the shorter wavelength. 
Using the same apparatus described here to produce a MgF beam, measurements of velocity as a function of arrival time show a similar velocity distribution for MgF and Mo.  The chirped MOT apparatus described here for trapping Mo atoms therefore appears promising for laser cooling and trapping MgF and many other light atoms and molecules.

\begin{acknowledgments}
The authors thank William McGehee, Qiaochu Guo, and Weston Tew for careful reading of the manuscript.  The authors thank Julia Scherschligt and Daniel Barker for technical assistance.  S. N. M. P. would like to
thank Stefan Truppe for useful discussions regarding
laser ablation. This work was funded by National Institute of Standards and Technology.
\end{acknowledgments}

\bibliography{references}
\end{document}